\documentclass[%
 aip,
 amsmath,amssymb,
 reprint,%
]{revtex4-1}

\usepackage{graphicx}
\usepackage{dcolumn}
\usepackage{bm}
\usepackage{xcolor}
\usepackage[utf8]{inputenc}
\usepackage[T1]{fontenc}
\usepackage{mathptmx}
\usepackage{etoolbox}
\usepackage[normalem]{ulem}

\makeatletter
\def\@email#1#2{%
 \endgroup
 \patchcmd{\titleblock@produce}
  {\frontmatter@RRAPformat}
  {\frontmatter@RRAPformat{\produce@RRAP{*#1\href{mailto:#2}{#2}}}\frontmatter@RRAPformat}
  {}{}
}%
\makeatother
\begin{document}

\preprint{AIP/123-QED}

\title[Noise-induced broadening of optical comb]{Noise-induced broadening of a quantum-dash laser optical frequency comb}
\author{A.I. Borodkin}
 \email{alex.i.borodkin@gmail.com}
\affiliation{ 
Université Côte d’Azur, Centre National de La Recherche Scientifique,\\ Institut de Physique de Nice, 06200 Nice, France
}%

\author{A.V. Kovalev}
\affiliation{%
ITMO University, St. Petersburg, 197101 Russia
}%

\author{M. Giudici}
\affiliation{
Université Côte d’Azur, Centre National de La Recherche Scientifique,\\ Institut de Physique de Nice, 06200 Nice, France
}

\author{G. Huyet}
\affiliation{
Université Côte d’Azur, Centre National de La Recherche Scientifique,\\ Institut de Physique de Nice, 06200 Nice, France
}

\author{A. Ramdane}
\affiliation{
Centre de Nanosciences et de Nanotechnologies, CNRS UMR 9001,\\Université Paris-Saclay, 91120 Palaiseau, France
}

\author{M. Marconi}
\affiliation{
Université Côte d’Azur, Centre National de La Recherche Scientifique,\\ Institut de Physique de Nice, 06200 Nice, France
}

\author{E.A. Viktorov}
\affiliation{%
ITMO University, St. Petersburg, 197101 Russia
}%

\date{\today}

\begin{abstract}
Single-section quantum dash semiconductor lasers have attracted much attention as an integrated and simple platform for the generation of THz-wide and flat optical frequency combs in the telecom C-band. In this work, we present an experimental method allowing to increase the spectral width of the laser comb by the injection of a broadband optical noise from an external semiconductor optical amplifier that is spectrally overlapped with the quantum dash laser comb. The noise injection induces an amplification of the side modes of the laser comb which acquire a fixed phase relationship with the central modes of the comb. We demonstrate a broadening of the laser comb  by a factor of two via this technique.
\end{abstract}

\maketitle

InAs/InP quantum dash (QDash) single-section laser diodes are unique in the family of the compact integrated mode-locked semiconductor lasers as they emit high quality optical frequency combs (OFC) with neither active nor passive modulation. For telecom or metrology applications, QDash OFC sources outperform similar devices in terms of timing jitter, amplitude and phase noise, and optical linewidth which can be as narrow as 15~kHz \cite{lelarge2007recent}. Nearly flat OFCs with about 10 nm bandwidth at -10 dB can be produced \cite{rosales2012high, duan2009high}, with tens of mW output power \cite{duan2009high,lelarge2007recent} and efficient power consumption \cite{van2006ultrafast}. These properties make single-section QDash lasers ideal for datacenter interconnects with an $\sim$~1~THz effective Quadrature Phase-Shift Keying (QPSK) bandwidth \cite{panapakkam2016amplitude}. By using an active electrical Radio Frequency (RF) modulation (active mode-locking), the effective bandwidth of the OFC source can be increased by more than 50 \% which is useful for applications  using QPSK and symbol rate of 12.5 GBd or higher \cite{panapakkam2016amplitude}. Besides, the active mode-locked operation generates OFCs with low $1/f$ noise and a corner frequency lower than 70 MHz,  which complies with the OFC standards for datacenter interconnects \cite{lelarge2007recent}.

\begin{figure}
\includegraphics[scale=0.4]{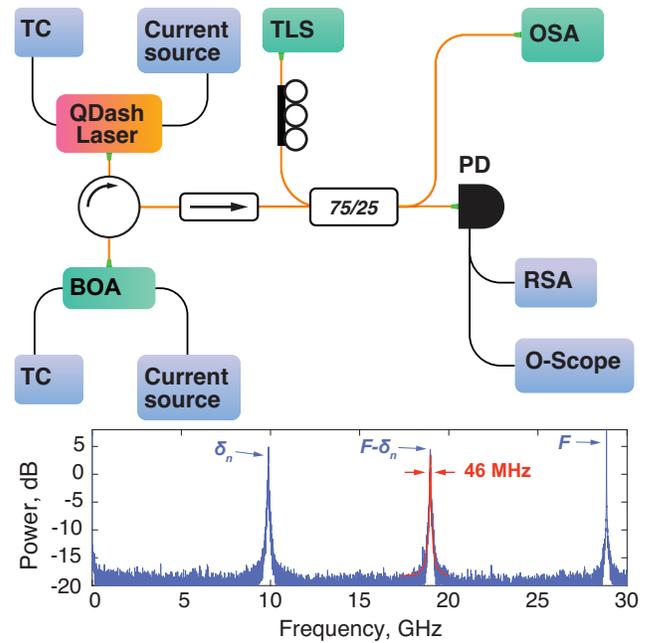}
\caption{\label{fig:setup} Experimental setup. BOA - Booster optical amplifier, TC~-~Temperature controller, TLS - Tunable laser source, PD - 35~GHz Photodetector, OSA - Optical spectrum analyser, O-scope - Oscilloscope, RSA - RF spectrum analyser. Inset: RF beat spectrum and Lorentzian fit (red), notations are described in the text.}
\end{figure}
The OFCs emitted by Qdash single section lasers display a linear phase chirp from -$\pi$ to $\pi$ across the whole spectrum and a group delay dispersion of a few $ps^2$ (refs\cite{rosales2012high,duill2016simple}). This generates a temporal output that is nearly CW . In these respects, the Qdash OFCs share similar properties with the Quantum Cascade Laser (QCL) \cite{singleton2018evidence} and Quantum dot laser combs \cite{hillbrand2020phase}. 
Recent theoretical works propose that the self-generated OFCs from single section Fabry-Pérot QDash lasers can be attributed to spatial hole burning and four-wave mixing \cite{dong2018physics,bardella2017self,opavcak2019theory}.The frequency-modulated (FM) comb formation in these systems obeys a variational principle \cite{piccardo2019frequency} which relies on the maximization of the total output power. That principle is responsible for the 2$\pi$ linear phase chirp across the FM comb spectra. 
\begin{figure*}
\includegraphics[scale=0.4]{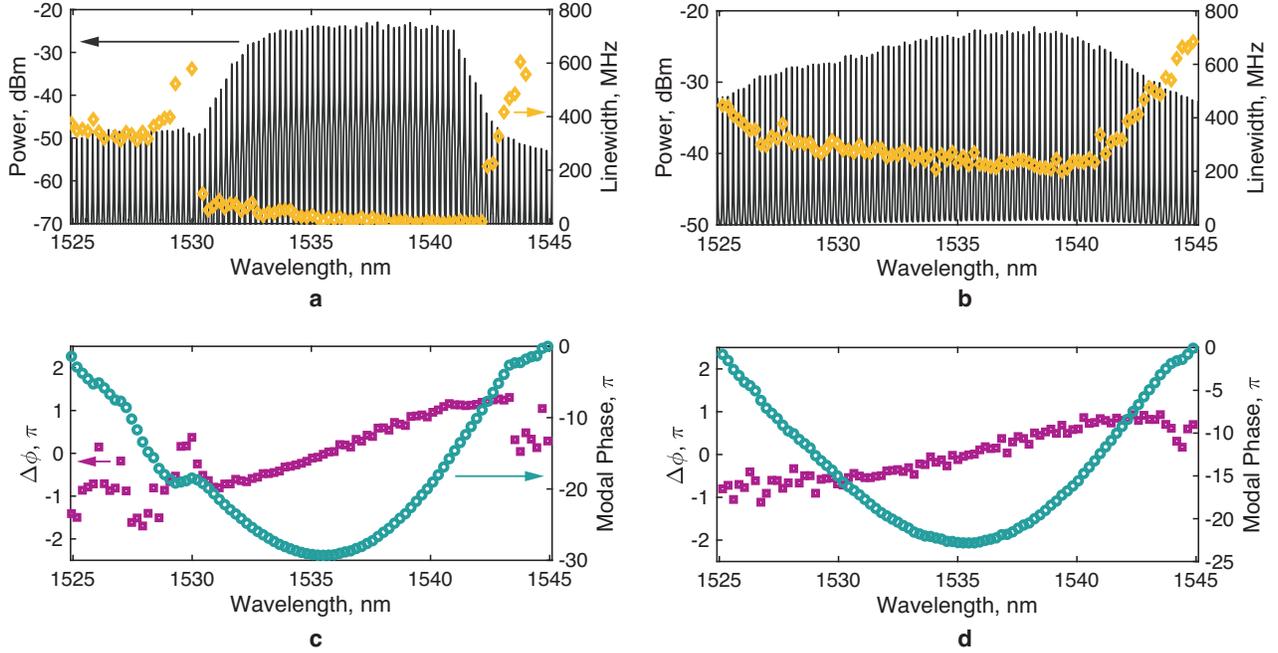}
\caption{Optical spectrum (black line), modal linewidth distribution (yellow markers),  spectral phase chirp (cyan markers), and modal phase (purple markers) of the QDash OFC laser: free running (a,c) and subject to an optical noise injection power of 2.2~mW (b,d).\label{fig:combs}}
\end{figure*}
The parabolic spectral phase of the FM combs emitted by the single section Qdash lasers can be compensated by propagating the laser output in a dispersion-compensation fiber. Nearly flat spectral phases can be obtained after propagation, which allows to generate 500~fs transform-limited pulses with RF repetition rates \cite{murdoch2011spectral}. \\
InAs/InP QDash materials are highly dense with strongly inhomogeneously broadened gain spectra \cite{somers2006optical,alizon2004multiple}, quantum wirelike properties of the states \cite{dery2004nature} and ps recovery times \cite{van2006ultrafast}.  Through a combination with two-photon absorption (TPA) these properties form an ultrafast gain response which was originally revealed in InAs/InP quantum dash amplifiers \cite{capua2010nearly} using multicolor pump probe spectroscopy technique \cite{wiesenfeld1979tunable, kesler1988subpicosecond}. Optimized InAs/InP structures allow high optical modal gain and ultra-fast mode-locked lasing from short cavities demonstrating 346~GHz pulse train with subpicosecond pulse durations \cite{merghem2009pulse}.


In this Letter, we demonstrate a substantial OFC broadening in a single-section  InAs/InP QDash laser subject to broadband optical noise injection. The broadening effect is strongly pronounced, and the OFC maintains the coherence between the modes, as was verified by a stepped-heterodyne (SH) measurement \cite{reid2010stepped}. We show that the parabolic shape of the spectral phase is preserved with noise injection and the group delay dispersion is reduced.  We link the effect of the OFC enhancement to a nearly instantaneous gain response \cite{capua2010nearly} which is unique to InAs/InP QDash gain media due to inhomogeneity of the gain broadening and quantum-wire-like density of states.\\
The experimental setup is shown in Figure \ref{fig:setup}. The InAs/InP QDash laser structures  contain 3~Qdash layers which provide sufficient gain for reducing the threshold current of the OFC generation down to only 25 mA. The cavity consists in a single section  1.5~mm long Fabry-Pérot with as-cleaved facets.  The free-running laser output consists in a nearly-flat OFC with 29~GHz Free Spectral Range (FSR), centered at 1530~nm, having a bandwidth of 12~nm at -10 dB and an average power of 10 mW at a pump power 5 times above threshold.
The 100~nm bandwidth optical noise injection is provided by a booster semiconductor optical amplifier (Thorlabs BOA1004P) via an optical fiber circulator and is spectrally centered at 1550~nm. In order to measure the spectral phase of the OFC, we realize the so-called "stepped-heterodyne" (SH) technique described in detail in \cite{reid2010stepped}. This technique consists in measuring the beatings between a  low-linewidth (400 kHz) tunable laser (TLS) source (Tunics 3642 HE CL) and the consecutive modes of the laser OFC. The beating signals between the modes $n$ and $n+1$ are multiplied with the complex conjugate of the beating signal at the FSR of the comb. This algorithm applied at each consecutive FSR allows to retrieve the phase relationship between the consecutive modes of the OFC. Such technique was already applied in \cite{verschelde2021analysis} to reconstruct the temporal envelop of the pulses emitted by a III-V-on-Si mode-locked laser.
 The beating signals between the OFC and the tunable laser are monitored using a fast photodiode (35 GHz) connected to a digital scope with 33 GHz bandwidth and 100 GS/s sampling rate. Part of this signal is also monitored with an optical spectrum analyzer (OSA, Yokogawa AQ6370D). 
An example of a  RF beating spectrum is shown in the inset of Figure \ref{fig:setup}, where $F$ is the FSR, $\delta_n$ and $F-\delta_n$ are the beat frequencies between the TLS and the $n^{th}$ and $(n-1)^{th}$ modes. From that measurement, we can directly infer the linewidth of the comb modes by applying a Lorentzian fit (solid red line) to the beating frequencies.

Figure \ref{fig:combs} shows the characteristics of the free-running (left column) and noise injected (right column) OFCs at 90~mA (3.6 times threshold) pump current with output power of 2~mW. The free running OFC acquired by the OSA (Fig.~\ref{fig:combs}a, black line) has a nearly flat profile and a 10~nm bandwidth (-10 dBm level). The much weaker side modes shown on the optical spectrum have an amplitude that is more than 20 dB lower than the flat central part of the OFC.
The central (side) modes have a linewidth of a few (hundreds) MHz, as indicated by the yellow markers. The phase difference ($\Delta\phi$) between consecutive modes (purple markers in Fig.~\ref{fig:combs}c,d) is obtained with the SH measurements. The phase chirp is linear for the flat part of the spectrum (from 1530~nm to 1542~nm) and covers the full range between $-\pi$ and $\pi$, as was previously reported for InAs/InP QDash laser \cite{rosales2012high}. The much weaker side modes show no fixed phase relationship with the central modes of the OFC.
The phase distribution (cyan markers in Fig.~\ref{fig:combs}c,d) is computed by integrating the phase difference and it displays a parabolic shape for the central modes of the OFC.
The effect of the broadband noise injection is shown in Fig.~\ref{fig:combs}b,d for a noise power of 2.2 mW. We observe from the optical spectrum (black line) that the amplitudes of the side modes are considerably increased, which induces an OFC broadening up to 20~nm (at -10~dB level). The linewidths of the central modes are increased (yellow markers) by about 2 orders of magnitude, while the linewidths of the side modes shows a gradual increase away from the OFC center. We observe from  Fig.~\ref{fig:combs}d that the linear phase chirp extends now over the integrality of the 20 nm width of the OFC and covers the full range between $-\pi$ and $\pi$. This indicates that the amplified side modes are naturally locked to the central modes of the OFC. As a consequence, the spectral phase distribution has acquired a parabolic shape (cyan markers in Fig.\ref{fig:combs}d) over the full spectrum.

\begin{figure}
\includegraphics[scale=0.4]{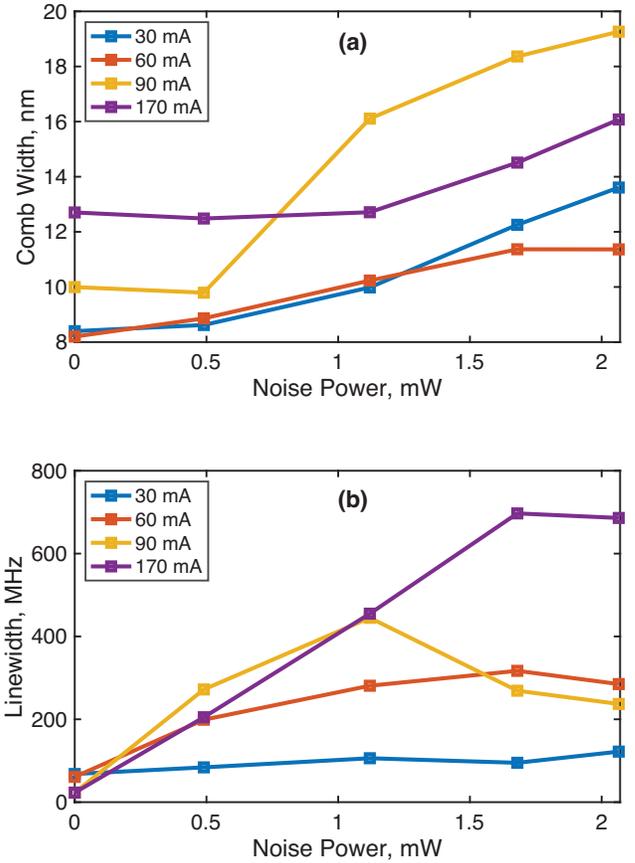}
\caption{\label{fig:widths} OFC width (a) and central mode linewidth (b) of the Qdash laser as a function of the noise power for different laser pump currents.}
\end{figure}

Figure \ref{fig:widths}a demonstrates that the increase of the OFC bandwidth and the locking of the side modes get more pronounced with the increase of the noise power in the whole range of the pumping current explored (up to $\sim$7 times threshold). In fact, we observe in Fig.~\ref{fig:widths}a that a low noise power (<~0.5~mW) does not affect the OFC width which sharply increases for intermediate 0.5~mW - 2~mW noise power. The linewidth of the central mode of the comb (Fig.~\ref{fig:widths}b) gradually increases with the noise power until a saturation is reached at about 1.5 mW. Despite the modal linewidth increase, the phase locking between the modes maintains over the whole range of injection power explored.

\begin{figure}
\includegraphics[scale=0.4]{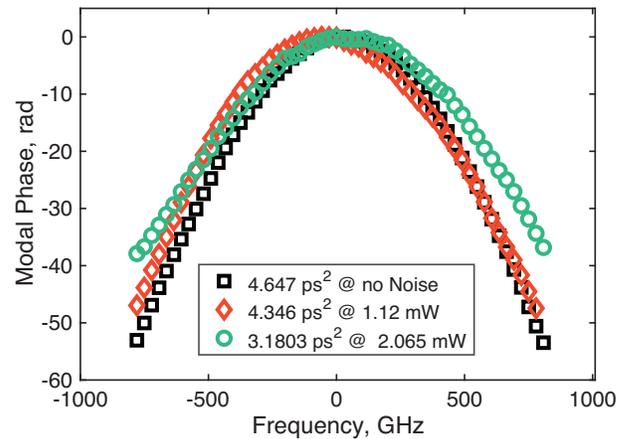}
\caption{\label{fig:disp} Spectral phase distribution of the Qdash laser OFC: free running (black squares) and subject to optical noise injection. The inset shows the group velocity dispersion values for two values of noise powers.}
\end{figure}

Highly-chirped mode-locked operation featured by the output frequency sweep from the red to blue edge of the OFC during one cavity round-trip time has previously been reported for single-section QDash lasers \cite{duill2016simple}. We used the measurements of the spectral phase in Figure \ref{fig:disp} to estimate the evolution of the group delay dispersion (GDD) with the increase of optical noise power based on the empirical relation from \cite{duill2016simple}. Each  distribution in Figure \ref{fig:disp} is well approximated by a parabola. We obtain from the parabolic fit a value of GDD equal to 4.647 ps$^2$ for the free running laser (black squares). This value is in agreement with the GDD reported in \cite{duill2016simple} for a similar laser. For noise powers of 1.12~mW and 2.065~mW , we obtain GDD values of 4.346 and 3.1803 $ps^2$ respectively. This demonstrates that increasing the noise level allows to decrease the GDD of the generated OFC.

We propose that the noise-induced OFC broadening relates to a phenomenon of nearly instantaneous gain process occurring in InAs/InP QDash gain media \cite{capua2010nearly}. The phenomenon has been revealed via multiwavelength pump probe measurements in InAs/InP QDash amplifiers and was explained due to the peculiarities of QDash material such as the inhomogeneous broadening of the gain and efficient nonlinear TPA \cite{capua2010nearly}. The injected high intensity optical noise ignites nonlinear absorption processes and excitation of the carriers to high energy levels. Ultrafast (10-100~fs) intraband carrier relaxation increases the ground state population and leads to nearly instantaneous increase of the QDash gain. It is followed by the OFC broadening. The effect of the nearly instantaneous gain response is not pronounced for low intensity optical pump and low pumping current similar to the low power noise experiments in Figure \ref{fig:widths}. The saturation of the spectral broadening at the high intensity noise (2 mW noise power) confirms the nonlinear character of the effect as the TPA process saturates with the increase of optical injection power.

In this work, we report a broadening of the OFC in a single section InAs/InP QDash laser caused by the injection of broadband optical noise. The broadening is due to an increase of the amplitude of the side modes which acquire a fixed phase relationship with the central modes of the combs. This effect is more pronounced when the noise level is increased, but it tends to saturate at a certain noise level. To our knowledge, it is the first experimental observation of this effect with OFC sources. The comb broadening, which is not achievable with pump power increase or signal amplification after the laser, can have useful applications for telecommunications and spectroscopy  applications. Moreover, the fact that all the amplified modes acquire a fixed phase relationship allows to potentially apply the same dispersion compensation to the emitted light as reported in ref\cite{rosales2012high} in order to generate ultrafast optical pulses.


This work has been supported by the French government, through the UCA-JEDI Investments in the Future project managed by the National Research Agency (ANR) with the reference number ANR-15-IDEX-01. This work was supported by the Ministry of Science and Higher Education of the Russian Federation, research project no. 2019-1442 (project reference number FSER-2020-0013).

\section*{Data Availability Statement}
The data that support the findings of this study are available from the corresponding author upon reasonable request.

\bibliography{refs}

\end{document}